\documentclass[longbibliography,aps,prr,twocolumn,superscriptaddress,showpacs,amsmath,amssymb, amsfonts,10pt]{revtex4-1}

\usepackage{amsmath}
\usepackage{amsfonts}
\usepackage{amssymb}
\usepackage{bm}
\usepackage{graphicx}
\usepackage{xcolor}
\usepackage[normalem]{ulem}

\usepackage[colorlinks,bookmarks=false,urlcolor=blue, citecolor=blue, linkcolor = blue]{hyperref}

\usepackage{enumerate}

\newcommand{\rr}{\mathbf{r}}

\newcommand{\dd}{\mathrm{d}}
\newcommand{\ee}{\mathrm{e}}
\newcommand{\kk}{\mathbf{k}}

\newcommand{\Svec}{\mathbf{S}}
\newcommand{\qq}{\mathbf{q}}

\newcommand{\mm}{\mathbf{m}}
\newcommand{\ff}{\bm{\phi}}

\begin{document}
	
	\title{Generalized Higgs mechanism in long-range interacting quantum systems}

	\author{Oriana K. Diessel}
	\affiliation{Max-Planck-Institute of Quantum Optics, Hans-Kopfermann-Strasse 1 , 85748 Garching, Germany}
	\affiliation{Munich Center for Quantum Science and Technology (MCQST), Schellingstr. 4, 80799 Munich, Germany}
	
	\author{Sebastian Diehl}
	\affiliation{Institute for Theoretical Physics, University of Cologne, Z\"{u}lpicher Strasse 77, 50937 Cologne, Germany}
	
	\author{Nicol\`{o} Defenu}
	\affiliation{Institute for Theoretical Physics, ETH Zurich, Wolfgang-Pauli-Strasse 27, 8049 Zurich, Switzerland}
	
	\author{Achim Rosch}
	\affiliation{Institute for Theoretical Physics, University of Cologne, Z\"{u}lpicher Strasse 77, 50937 Cologne, Germany}
	
	\author{Alessio Chiocchetta}
	\affiliation{Institute for Theoretical Physics, University of Cologne, Z\"{u}lpicher Strasse 77, 50937 Cologne, Germany}

	\begin{abstract}
		The physics of long-range interacting quantum systems is currently living through a renaissance driven by the fast progress in quantum simulators. In these systems many paradigms of statistical physics do not apply and also the universal long-wavelength physics gets substantially modified by the presence of long-ranged forces. Here we explore the low-energy excitations of several long-range interacting quantum systems, including spin models and interacting Bose gases, in the ordered phase associated with the spontaneous breaking of U(1) and SU(2) symmetries. Instead of the expected Goldstone modes, we find three qualitatively different regimes, depending on the range of the interaction. In one of these regimes the Goldstone modes are gapped, via a generalization of the Higgs mechanism. Moreover, we show how this effect is realized in current experiments with ultracold atomic gases in optical cavities.   
	\end{abstract}
	
	\date{\today}
	
	\maketitle
	
	Many-body systems with long-range interactions represent one of the most intriguing challenges in modern condensed matter, AMO and statistical physics. The long-range nature of the interactions, in fact, dispense these systems from some of the fundamental paradigms of statistical physics.
	
	For instance, classical long-range interacting systems (LRIS), such as gravitational systems and non-neutral plasmas, enjoy unusual properties, such as non-additivity of the energy and non-ergodicity. The latter may, in turn, lead to a very slow (if not completely absent) thermalization dynamics~\cite{Dauxois_2002,Campa_2009,Campa_2014}. Furthermore, the Mermin-Wagner theorem does not apply to LRIS, enabling them to exhibit spontaneous symmetry breaking even at low spatial dimensions~\cite{halperin2019hohenberg}. 
	
	Recently, the investigation of LRIS has gained new momentum from a flurry of experimental realizations of \emph{quantum} long range interacting systems (QLRIS), including Rydberg atoms~\cite{Saffman_2010}, dipolar quantum gases~\cite{Lahaye_2009}, polar molecules~\cite{Carr_2009}, quantum gases coupled
	to optical cavities~\cite{Mivehvar_2021,Ritsch2013}, trapped ions~\cite{Monroe_2021}, and dipolar magnets~\cite{castelnovo2008magnetic,DeBell2000}. 
	This increasing amount of experimental evidence opens up new challenges to theoretically understand the corresponding many-body problem. 
	
	A great deal of information on a quantum system is typically encoded in its~\emph{low-energy properties}. 
	Remarkably, while most of the  existing theoretical studies involve the characterization of ground states and critical properties of QLRIS~\cite{Defenu_2021}, little is known about their low-energy spectrum, with few remarkable exceptions, including their point-spectrum nature~\cite{Defenu_2021_pnas}, the existence of confinement \cite{Liu_2019}, and of fractional excitations~\cite{Chiocchetta2021,Birnkammer_2020}.
	Among these, it was pointed out that long-range interactions may cause the Goldstone modes to acquire a mass, thus violating the Goldstone theorem. Notable examples of this include a Bose gas with Coulomb interaction in 3+1 dimensions~\cite{Foldy1961}, a superconductor with Coulomb interactions in 3+1 dimensions~\cite{Anderson1963}, and the Schwinger model in 1+1 dimension (called in this context \emph{seizing of the vacuum})~\cite{Kogut_1975}. Even the celebrated Higgs mechanism appears to be a special case of this more general mechanism, as the gauge fields effectively mediate a long-range Coulomb interaction~\cite{Strocchi1986, Strocchi2008}.
	
	Our study targets the unexplored connection between the expanding field of QLRIS and the Higgs mechanism, the latter being at the center of increasing experimental attention, e.g., in strongly interacting Fermi superfluids~\cite{behrle2018higgs}, including pioneering studies of its few-body precursor\,\cite{bjerlin2016few,bahya2020observing}. The textbook Higgs mechanism can be understood from two angles: 
	\begin{figure*}[t!]
		\includegraphics[width=\textwidth]{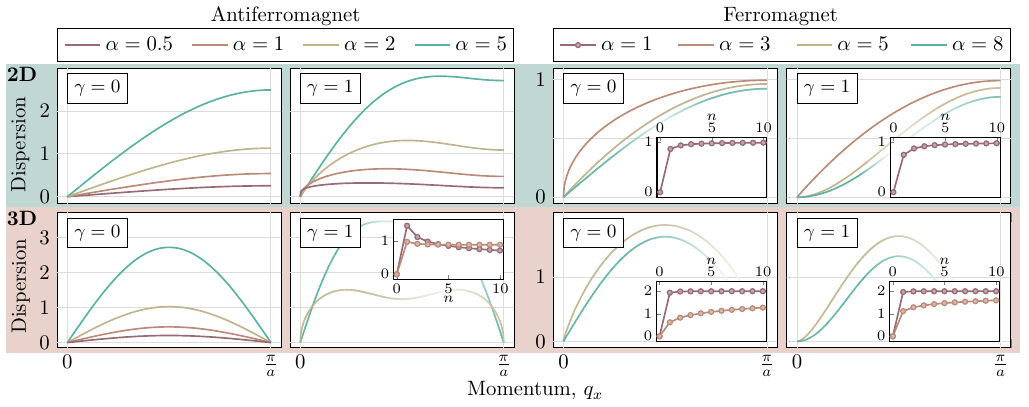}
		\caption{Two-dimensional (upper green row) and  three-dimensional (lower red row) spin-wave spectrum $E_\qq$ as a function of $\qq = (q_x,0)$ ($\qq = (q_x,0,0)$) from Eq.~\eqref{eq:Heisenberg_Hamiltonian} for an AFM (left) and FM (right) interaction. The insets show the rescaled discrete spectrum $E_\mathbf{n}$ as a function of the quantum number $\mathbf{n} = (n_x,0)$ ($\mathbf{n} = (n_x,0,0))$.  The unrescaled dispersions which diverge with the system size are shown in App.~\ref{app_c}, Fig.~\ref{fig:DispersionsSM}. All results are obtained for linear system sizes $L=800$.}
		\label{fig:Dispersions}
	\end{figure*}
	on the one hand, in a gauge theory the Goldstone mode can be absorbed by a gauge transformation into the electromagnetic field which then obtains a mass by coupling to the condensate. On the other hand, the 
	integration of the gapless electromagnetic field gives rise to long-range interactions, which also provide a mass for the Goldstone mode.
	Here, we uncover a new side of this second line of thought, by showing the occurrence of a \emph{generalized Higgs mechanism} in QLRIS for different kinds of long-range interactions.
	
	We consider a number of experimentally-relevant models,  including anti- and ferromagnetic spin models, and weakly interacting Bose gases, which exhibit spontaneous breaking of continuous symmetries, and analyze their low-energy excitation spectrum for different spatial dimensions and very general long-range interactions. 
	
	Our first main result is to show for a broad range of models that three qualitatively different regimes exist depending on the interaction range:
	I. a regime where the Goldstone modes behave as in short-range models, II. a regime where the Goldstone dispersion is gapless but qualitatively different at low momenta, and III. a regime where the Goldstone modes are gapped.
	Our second main result is the characterization of the visibility of the generalized Higgs mechanism in current experiments with atomic gases in optical cavities: we demonstrate that, with realistic experimental parameters, the Bogoliubov dispersions are gapped and discrete, thus realizing the third regime.
	
	\emph{Antiferromagnetic Heisenberg model---}
	The anisotropic antiferromagnetic Heisenberg model on a square lattice with long-ranged interactions is given by:
	\begin{equation}
		\label{eq:Heisenberg_Hamiltonian}
		\hat{H} = \sum_{ i \neq j }\, J_{ij} (\hat{S}^x_i\hat{S}^x_j + \hat{S}^y_i\hat{S}^y_j + \gamma \hat{S}^z_i\hat{S}^z_j),
	\end{equation}
	where $\hat{S}^x_i, \hat{S}^y_i, \hat{S}^z_i$ are spin operators of length $S$ residing at the $i-$th lattice site, $0 \leq \gamma \leq 1 $ is the anisotropy coefficient, and $J_{ij} > 0 $ is a long-range antiferromagnetic exchange, which we will assume to be 
	$J_{ij} = |\rr_{ij}|^{-\alpha}$,
	with $\alpha > 0$ and $\rr_{ij} \equiv \rr_i-\rr_j$ the relative distance between the $i$-th and $j$-th sites. Since  $J_{ij}$ couples  all the spins antiferromagnetically, it acts similarly to a frustrating interaction~\cite{Chiocchetta2021}.
	We will focus on the case of semi-classical spins (i.e., with spin length $S\gg 1$), for which the spin-wave approximation and non-linear sigma model analyses provide accurate results. The case for $S=1/2$ was considered in Refs.~\cite{Birnkammer_2020} and~\cite{Chiocchetta2021} for the one- and two-dimensional case, respectively, and it was shown to lead to exotic phases including quantum spin liquids and valence-bond solids.\\
	\begin{figure*}[t!]
		\includegraphics[width=\textwidth]{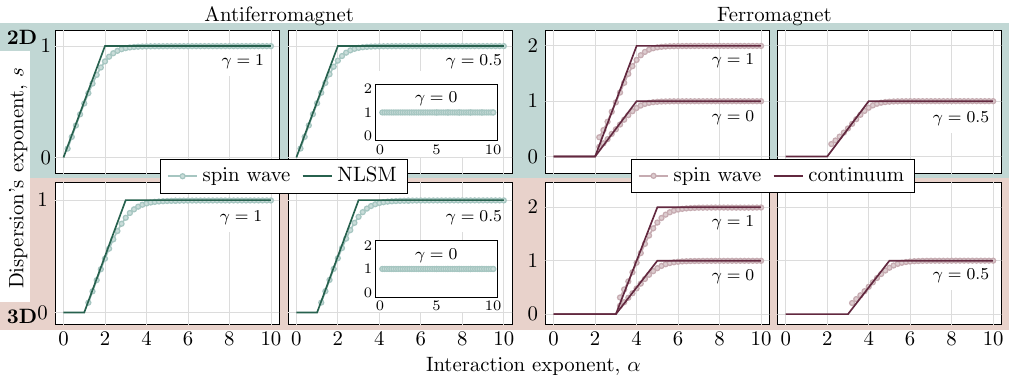}
		\caption{Low-momentum dispersion exponent $s$ as a function of the interaction exponent $\alpha$. Results for $d=2$ are shown in the upper green row, for $d=3$ in the lower red row, for an antiferromagnetic interaction on the left side (green curves), and for a ferromagnetic interaction on the right side (red curves). 
			The dots are obtained by fitting the results of the spin-wave analysis via $E^{(+)}_\qq\propto |\qq|^s$, while the solid lines show the analytical results of the NLSM analysis/continuum-limit approximation.
			All results from the spin wave analysis are obtained for linear system sizes $L=800$.}
		\label{fig:Exponent}
	\end{figure*}
	The classical (i.e., for $S\to \infty$) ground state of~\eqref{eq:Heisenberg_Hamiltonian} is a N\'{e}el state~\cite{Chiocchetta2021}, which allows one to define two sublattices on each of which the classical spin points along the same (and opposite) direction.  In order to account for quantum fluctuations around this state, a Holstein-Primakoff transformation is used, and the Hamiltonian is truncated at order $O(S)$~\cite{Altland_2010} (see App.~\ref{App:1}). By diagonalising the resulting Hamiltonian with a Bogoliubov transformation, one  finds the spectrum of the low-energy excitations (i.e., spin waves), given by $
	E^{\pm}_\qq = S \sqrt{ (J^\text{d}_0 \mp J^\text{d}_\qq \!-\!J^\text{s}_0 + J^\text{s}_\qq)[J^\text{d}_0 - J^\text{s}_0 +\gamma(J^\text{s}_\qq \pm J^\text{d}_\qq)]}$,
	with $\qq$ a quasi-momentum in the first Brillouin zone of the sublattice, and $J^\text{s}_\qq \equiv \sum_{\ell \in \text{same}} \ee^{-i \qq\cdot \rr_\ell} J_\ell $ and $  J^\text{d} _\qq \equiv \sum_{\ell \in \text{diff} } \ee^{-i \qq\cdot \rr_\ell} J_\ell,$ where the first (second) sum includes only vectors connecting sites on the same (different) sublattice.
	The dispersion $E^{+}_\qq$ corresponds to the Goldstone excitation branch, while $E^{-}_\qq$ corresponds to generally gapped excitations. $E^{+}_\qq$ is shown in Fig. \ref{fig:Dispersions} in the left panel for different values of $\gamma$ and $\alpha$, and for different spatial dimensions. In the SU(2)-symmetric case ($\gamma=1$), the two dispersions become degenerate and both gapless, as two Goldstone modes are expected for the spontaneous breaking of a SU(2) symmetry. We numerically evaluated the dispersions, and analyse the low-momenta behaviour of the Goldstone branch by parametrizing the dispersion as $E^{+}_\qq \approx A |\qq|^s$, for $\qq\to 0$, and use a fit to determine the value of $s$ as a function of the exponent $\alpha$ in the Hamiltonian~\eqref{eq:Heisenberg_Hamiltonian} (see Fig.~\ref{fig:Exponent}).  For nearest-neighbour interactions, one expects a linear spectrum~\cite{Altland_2010}, i.e.,  $s=1$. A gapped dispersion, instead, would correspond to $s=0$.
	
	Before discussing the results, it is convenient to get a fully analytical expression for $s(\alpha)$. To this end, we use the non-linear sigma model (NLSM), which is able to capture the low-energy behaviour of an antiferromagnet in the SU(2)-symmetric case~\cite{auerbachbook} (i.e., $\gamma = 1$). We represent a classical spin of length $S$ as $\Svec_j/S = (-1)^j(1-\mm_{j}^{2})^{1/2} \bm{\phi}_{j}+\mm_{j}, $ where $\ff_j$ describes the order parameter associated with the N\'{e}el state, $\mm_j$ is the canting field, and they satisfy the conditions $\boldsymbol{m}_{j} \cdot \boldsymbol{\phi}_{j}=0$ and $\left|\boldsymbol{\phi}_{j}\right|^{2}=1$, as a consequence of the spin-length conservation.
	Assuming the canting fluctuations to be small, i.e.,  $\mm_j^2 \ll 1$, and by performing a spatial coarse-graining, the effective action $\mathcal{A}$ of the Heisenberg model reads~\cite{auerbachbook} (see App.~\ref{App:2}):
	\begin{equation}
		\label{eq:action_2}
		\mathcal{A} = \! \int_{t,\rr} \!\!\left[ \boldsymbol{m}_\rr \cdot(\dot{\boldsymbol{\phi}_\rr} \!\times\! \boldsymbol{\phi}_\rr)
		\!-\! (\nabla \ff_\rr)^2 - \!\!\int_{\rr'} \!\!\!J_{\rr-\rr'}\, \mm_\rr\!\cdot\! \mm_{\rr'}\! \right]\!\!,
	\end{equation}
	with $\int_{t,\rr} \equiv \int \dd t\,  \dd^d r$, and $J_\rr$ the continuum version of $J_{ij}$. Note that the long-ranged interaction is only activated by the canting field. 
	To analyse the excitation spectrum of this effective theory, we expand around a homogeneous order parameter $\ff_0$ as $\ff_\rr = \ff_0 + \delta\ff_\rr$, and only retain terms quadratic in $\delta\ff_\rr$ and $\mm_\rr$. The resulting quadratic action can be diagonalized, and leads
	to the quasiparticle dispersion $E_\qq \propto \sqrt{|\qq|^2 J_\qq}$. For $\alpha>d-2$, with $d$ the spatial dimension, the low-momentum behaviour of $E_\qq$ reads 
	\begin{equation}
		\label{eq:dispersion_AFM}
		E_\qq \approx 
		\begin{cases}
			|\qq| & \text{for } \alpha > d \\
			|\qq| \sqrt{\log|\qq|} & \text{for } \alpha = d\\
			|\qq|^\frac{2+\alpha - d}{2}  & \text{for } d-2<\alpha < d.
		\end{cases}
	\end{equation}
	For $\alpha\leq d-2$, the Fourier transform of $J_\rr$ diverges with the system size, and therefore a regularization is needed. This is achieved by  rescaling the proper timescales for excitation propagation, in analogy with the well-known case of diverging ferromagnetic interactions\,\cite{antoni1995clustering,bachelard2013universal}. This procedure is outlined in App.~\ref{app_c}, and reveals the discrete, gapped nature of the spectrum, see insets in Fig.~\ref{fig:Dispersions}. Once properly regularized, it becomes evident that the divergent dispersion relation characteristic of the third regime reduces to a pure point spectrum, similar to the one observed in strongly disordered systems, which, as pointed out in Ref.~\cite{Defenu_2021_pnas}, leads to the breakdown of conventional equilibration and irreversibility concepts. This phenomenon, already described in Ref.\,\cite{Defenu_2021_pnas} for ferromagnetic systems, is found  here to also occur in antiferromagnetic systems with $\alpha\leq d-2$. 
	
	The case $\gamma<1$ can be also treated within the NLSM~\cite{Haldane1983} (see App.~\ref{App:2}), and it renders the same result for $s(\alpha)$, except for the case $\gamma=0$, where it predicts a linear dispersion, regardless of the interaction. 
	\begin{figure}[h!]
		\includegraphics[width=0.48\textwidth]{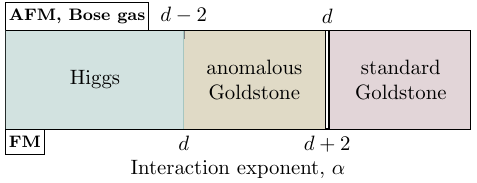}
		\caption{Regimes for the long-range-interacting FM and AFM  Heisenberg model (for $\gamma>0$) and a Bose gas as a function of the interaction exponent $\alpha$. The double vertical line represent the presence of logarithmic corrections (cf. Eqs.~\eqref{eq:dispersion_AFM},\eqref{eq:dispersion_FM_1})}
		\label{fig:PhaseDiagram}
	\end{figure}
	The resulting values of $s(\alpha)$ are in agreement with the values obtained from the spin-wave approximation, as shown in Fig.~\ref{fig:Exponent}, with minor discrepancies due to finite-size effects. Note that for $\alpha \leq d-2$ we did not fit the values of $s(\alpha)$, as the spectrum is discretized.
	
	This is the first main result of this work: three qualitatively different regime exist for the Goldstone mode, depending on the value of $\alpha$.
	\begin{enumerate}[I.]
		\item The Goldstone mode is as in the short-range model.
		\item The Goldstone mode is anomalous.
		\item The Goldstone mode is gapped and discrete.
	\end{enumerate}
	In particular, the last regime hosts the \emph{generalized Higgs mechanism}: for sufficiently long-ranged interactions, the Goldstone spectrum is discrete and, in turn, it becomes gapped .
	An instance of regime II. was found for a dipolar ($\alpha=3$) antiferromagnet on a square lattice~\cite{Peter_2012}. 
	
	From Fig.~\ref{fig:Exponent} we also observe that, for $0<\gamma\leq 1$, the same function $s(\alpha)$ holds, while for $\gamma=0$, i.e., for the quantum $XY$ model, $s(\alpha)$  is independent of $\alpha$ and equals 1, in agreement with the NLSM. 
	We also emphasize here that Goldstone modes can acquire a mass when a symmetry-breaking perturbation is added to the system. In general, the determination of this mass is  non-trivial  and is currently subject of active research~\cite{Watanabe_2013}.

	\emph{Ferromagnetic Heisenberg model---}
	We consider now an anisotropic  ferromagnetic Heisenberg model on a square lattice, which takes the same form as Eq. \eqref{eq:Heisenberg_Hamiltonian}, with long-ranged interactions $J_{ij} =-|\rr_{ij}|^{-\alpha}$.
	The ground state of this model is an ordinary ferromagnet, and the low-energy excitations can again be derived using the spin-wave analysis. The Holstein-Primakoff transformation leads to the spin-wave spectrum $E_\qq=S\sqrt{(J_0-J_\qq)(J_0-\gamma J_\qq)}$ where $J_\qq \equiv \sum_{\ell} \ee^{-i \qq\cdot \rr_\ell} J_\ell.$ The low-momentum behaviour of $E_\qq$ can be easily derived analytically by approximating the long-range interaction as $J_\qq \approx \int_{|\rr|>a} \dd^d \rr |\rr|^{-\alpha} \ee^{-i\qq\cdot \rr}$, with $a$ the lattice spacing, and for $\alpha > d$ is given by
	\begin{equation}
		\label{eq:dispersion_FM_1}
		E_\qq \approx 
		\begin{cases}
			|\qq|^{2x_{\gamma}} & \text{for } \alpha > d+2 \\
			(|\qq|^{2} \log|\qq|)^{x_{\gamma}} & \text{for } \alpha = d+2\\
			|\qq|^{(\alpha - d)x_{\gamma}}  & \text{for } d<\alpha < d+2
		\end{cases}
	\end{equation}
	with $x_{\gamma}=1$ for $\gamma=1$ and $x_{\gamma}=1/2$ for $\gamma<1$. 
	For $\alpha \leq d$, the dispersion diverges with the system size and a regularization is therefore added as for the AFM (see App.~\ref{app_c}).
	The dispersions $E_\qq$ for different values of $\alpha$, $\gamma$, and spatial dimensions are also shown in Fig. \ref{fig:Dispersions}, right panel. The existence of a gapped Goldstone mode in the ferromagnetic Heisenberg model due to long-ranged interactions was first pointed out in Refs.~\cite{Lange_1965,Lange_1967}.
	
	In Fig.~\ref{fig:Exponent} we show the curves $s(\alpha)$ in comparison with the fit obtained by the lattice evaluation of the spin-wave dispersion, showing good agreement.
	The results show again, as in the AFM case, the existence of the same three different regimes for $s(\alpha)$ (cf. Fig.~\ref{fig:Exponent}).
	Differently from the AFM case, the result is sensitive to the symmetry of the model, namely for the SU(2) ($\gamma=1$) and the U(1) ($\gamma < 1$) cases. While for both cases the regimes boundaries are the same, the values of the exponents change. 
	Instances of regime II were found in several FM models: 
	a ferromagnetic XXZ chain~\cite{Vanderstraeten2018} ($d=1, \gamma<1, \alpha>1$ ), a ferromagnetic U(1)-symmetric spin system with dipolar interactions on a square lattice~\cite{Maalev1976,Peter_2012} ($d=2$, $\gamma<1$, $\alpha=3$), and a ferromagnetic SU(2)-symmetric spin system~\cite{Buchheit2022} ($d$ generic,$\gamma=1$, $\alpha>d$).
	
	It is worth noting that both FM (with $\gamma<1$) and  AFM interactions yield the same scaling $s(\alpha)$, but in different range of the interaction exponent $\alpha$, such that $s_{\rm FM}(\alpha)=s_{\rm AFM}(\alpha-2)$, see Fig.\,\ref{fig:PhaseDiagram}. The same correspondence has been observed between the critical exponents of ferromagnetic rotor models and antiferromagnetic spin Hamiltonians\,\cite{defenu2017criticality} and we conjecture it to constitute a generic feature of long-range interactions.

	\emph{Interacting Bose gas---}
	The results shown above apply in a similar fashion to the case of a condensed Bose gas with long-range interactions. This was first studied in the context of a charged Bose gas, where the particles interact via a Coulomb potential. There, it was shown that the Bogoliubov spectrum is gapped in three dimensions~\cite{Foldy1961}. We consider in the following a generalization of this model,  relevant for ongoing experiments with cold-atoms in a cavity~\cite{Lev2018}.
	\begin{figure}[h!]
		\includegraphics[width=0.47\textwidth]{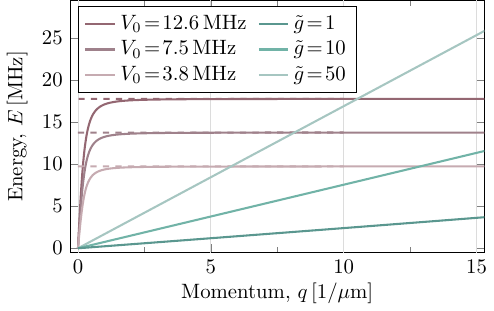}
		\caption{Bogoliubov dispersions for a quasi-two dimensional Rubidium gas, for different interaction potentials $V_\rr = V_0 K_0(Q|\rr|)$ (solid red curves), logarithmic (dashed red curves), and contact interaction (solid green curves).}
		\label{fig:Experiment}
	\end{figure}
	We assume the Hamiltonian of the gas to be given by 
	\begin{equation} 
		H =\int_\rr\left( -\psi_\rr^\dagger\frac{\nabla^2}{2m} \psi_\rr  +\frac{1}{2}  \int_{\rr'}  V_{\rr^{\prime}-\rr} \psi^\dagger_\rr\psi^\dagger_{\rr'}\psi_{\rr'}\psi_\rr\right),
	\end{equation}
	with a long-ranged interaction $V_\rr = V_0|\rr|^{-\alpha}$.
	According to Bogoliubov's theory, the bosonic field can be decomposed in a homogeneous condensate and fluctuations around it, i.e., $\psi_\rr = \psi_0 + \bar{\psi}_\rr$. By replacing it in the Hamiltonian, retaining terms up to quadratic order in the fluctuations, and finally diagonalizing via a Bogoliubov transformation, one obtains
	$E_\qq = \sqrt{\epsilon_\qq (\epsilon_\qq + 2n_0V_\qq )}$
	with $\epsilon_\qq = \hbar^2 |\qq|^2/2m$ the free particle dispersion, $n_0=|\psi_0|^2$ the condensate density, and $V_\qq$ the Fourier transform of the interaction potential. The chemical potential is set to $\mu = n_0 V_{\qq=0}$ for thermodynamical stability. %
	By expanding $E_\qq$ at low momenta, we find the function $s(\alpha)$, given in Fig. \ref{fig:PhaseDiagram}, which is the same as for the case of the AFM Heisenberg model with $\gamma>0$. Correspondingly, the same three regimes, depending on the value of $\alpha$, can be recognized.  
	
	\emph{Visibility in ultracold quantum gases---}
	We consider the following simplified model, inspired by the experimental setup of Ref.~\onlinecite{Lev2018}, consisting of a quasi-two-dimensional gas of bosonic  $^{87}$Rb atoms enclosed in a multimode cavity.
	The atomic interaction mediated by the cavity modes has the form (far from the trap boundaries) $V_\rr = V_0 K_0(Q|\rr|)$
	where $V_0$ can be varied upon tuning the pump Rabi frequency and $Q \simeq 0.29 \mu\text{m}^{-1}$ depends on the number of modes coupled to the atoms via the cavity, and on the cavity mode waist.
	The modified Bessel function $K_0(x)$ falls off exponentially for $x\gg 1$, while $K_0(x) \approx -\log x$ for $x\ll 1$. Accordingly, we expect the deviation from linearity and the consequent gapping of the Bogoliubov dispersion to be visible for momenta $|\qq| > Q$.
	In fact, the corresponding Bogoliubov dispersion reads $ E_\qq = \sqrt{\epsilon_\qq \left(\epsilon_\qq +  2 n_0 V_0/(|\qq|^2 + Q^2 ) \right)},$
	which shows a plateau for $|\qq| > Q$, corresponding to the gap one would have in the pure long-range case with $Q=0$.
	To better grasp the visibility of this effect in said experimental platform, we report in Fig.~\ref{fig:Experiment} the dispersions $E_\qq$ for experimental values of $n_0 = 5.5 \times 10^3 (\mu\text{m})^{-2}$ and $m = 87 m_p$ (with $m_p$ the proton mass), for different values of $V_0$ (solid red lines), and compare them with the corresponding dispersions with $Q=0$ (dashed red lines). The dispersions with finite $Q$ substantially overlap with the gapped ones with $Q=0$. To emphasize the difference with the dispersion for usual contact interactions, we additionally plot the Bogoliubov dispersion for $V(\rr) = \tilde{g} (\hbar^2/2m) \delta^{(2)}(\rr)$, with different values of the dimensionless parameter $\tilde{g}$ (solid green lines). The difference with the dispersions for the long-range interacting model is evident, for the range of parameters used. This shows that the generalized Higgs mechanism can be observed in current experimental setups. For a quantum gas in a cavity, the excitations can be probed, e.g., using Bragg spectroscopy~\cite{Mottl_2012}.
	
	\emph{Conclusions}--- We showed that, for a number of experimentally-relevant quantum systems, Goldstone modes can be strongly affected by the presence of long-range interactions. In comparison to short-range interacting systems, the Goldstone dispersion can be distorted or, even more remarkably, be gapped, in what can be regarded as a generalized Higgs mechanism. We showed that this mechanism can be detected in current experiments with ultracold atomic gases in an optical cavity. Our results open intriguing perspectives for the engineering of quantum materials. For example, an experimentally tunable interaction range would enable the switching between gapped und gapless spectra, resulting in very different thermodynamical properties.

	\emph{Acknowledgments}--- We warmly acknowledge discussions with  J.~Keeling,  B.~Lev and  A.~Trombettoni. We are especially thankful to D.~Kiese for carefully reading the manuscript and providing helpful comments.  We acknowledge support from the European Research Council (ERC) under the Horizon 2020 research and innovation program, grant agreement No. 647434 (DOQS), by the Deutsche Forschungsgemeinschaft (DFG, German Research Foundation)  CRC 1238 project number 277146847, CRC 1225 (ISOQUANT) project number 273811115, and under Germany’s Excellence Strategy 
	EXC2181/1-390900948 (the Heidelberg STRUCTURES Excellence Cluster).
	O.~K.~D. acknowledges funding from the International Max Planck Research School for Quantum Science and Technology (IMPRS - QST). 
	
	\bibliography{bibliography.bib}

\appendix
\section{Spin wave analysis
\label{App:1}}
In this Appendix, we give more details on the spin wave analysis of Hamiltonian~\ref{eq:Heisenberg_Hamiltonian}. We do a large $S$ expansion around the classical groundstate, which is either the Néel state or the ferromagnetic state, depending on the form of the interaction. 
The first case calls for the definition of two sublattices, $A$ and $B$, on each of which the spins point in the same direction.
Expressing the spin vectors by ladder operators $S^+=S^x+iS^y$ and $S^-=S^x-iS^y$ allows to apply the Holstein-Primakoff transformation  
\begin{subequations}
	\label{eq:HP}
	\begin{align}
		S^{z}_i&=S-a_i^{\dagger} a_i, \\ 
		S_i^{+}&=\big(2 S-a_i^{\dagger} a_i\big)^{1 / 2} a_i, \\
		S_i^{-}&=a_i^\dagger \big(2 S-a_i^{\dagger} a_i\big)^{1 / 2},
	\end{align}
\end{subequations}
\\

\noindent for $i \in A$, with $a_i$ and $a_i^\dagger$ bosonic operators, and
\begin{subequations}
	\begin{align}
		S^{z}_i&=b_i^{\dagger} b_i-S, \\ 
		S_i^{+}&=b_i^\dagger\big(2 S-b_i^{\dagger} b_i\big)^{1 / 2} , \\
		S_i^{-}&=\big(2 S-b_i^{\dagger} b_i\big)^{1 / 2} b_i,
	\end{align}
\end{subequations}
for $i \in B$, with $b_i$ and $b_i^\dagger$ bosonic operators.
By replacing these transformations in Eq.~\ref{eq:Heisenberg_Hamiltonian} and retaining only terms up to order $S$, one obtains a Hamiltonian quadratic in the bosonic operators, which can then be Fourier transformed via
$a_j^{\dagger}=N^{-1/2}\sum_\kk a_{\kk}^{\dagger}\ee^{i\kk\rr_j}$, with $N$ the system volume,
into quasi-momentum space. The eigenvalues can be obtained by diagonalizing the matrix

\begin{widetext}
	\begin{align}
		\label{Eq:AFM_Matrix}
		i\frac{d}{dt}
		\begin{pmatrix}
			a_{\qq}\\
			a^{\dagger}_{-\qq}\\
			b_{\qq}\\
			b^{\dagger}_{-\qq}
		\end{pmatrix}
		=
		\frac{S}{2}
		\begin{pmatrix}
			2J^{\text{d}}_0-2J^{\text{s}}_0+\gamma^+J^{\text{s}}_{\qq} &\gamma^-J^{\text{s}}_{-\qq} & \gamma^-J^{\text{d}}_{\qq} &\gamma^+J^{\text{d}}_{-\qq}\\
			-\gamma^-J^{\text{s}}_{-\qq} & -2J^{\text{d}}_0+2J^{\text{s}}_0-\gamma^+J^{\text{s}}_{\qq} & -\gamma^+J^{\text{d}}_{-\qq} &-\gamma^-J^{\text{d}}_{\qq}\\
			\gamma^-J^{\text{d}}_{\qq}& \gamma^+J^{\text{d}}_{-\qq} & 2J^{\text{d}}_0-2J^{\text{s}}_0+\gamma^+J^{\text{s}}_{\qq}&\gamma^+J^{\text{d}}_{-\qq}\\
			-\gamma^+J^{\text{d}}_{-\qq} & -\gamma^-J^{\text{d}}_{\qq} & -\gamma^-J^{\text{s}}_{-\qq}&-2J^{\text{d}}_0+2J^{\text{s}}_0-\gamma^+J^{\text{s}}_{\qq}
		\end{pmatrix}
		\!\!\!
		\begin{pmatrix}
			a_{\qq}\\
			a^{\dagger}_{-\qq}\\
			b_{\qq}\\
			b^{\dagger}_{-\qq}
		\end{pmatrix}
	\end{align}
\end{widetext}

in the antiferromagnetic, and

\begin{align}
	\label{Eq:FM_Matrix}
	\!\!\!\!\!\!\!\!i\frac{d}{dt}\!
	\begin{pmatrix}
		a_{\qq}\\
		a^{\dagger}_{-\qq}
	\end{pmatrix}
	\!=\!
	S\!
	\begin{pmatrix}
		2J_0\!-\!\gamma^+J_{-\qq} &-\gamma^-J_{\qq} \\
		\gamma^-J_{-\qq} & -2J_0\!+\!\gamma^+J_{\qq} 
	\end{pmatrix}
	\!\!\!
	\begin{pmatrix}
		a_{\qq}\\
		a^{\dagger}_{-\qq}
	\end{pmatrix}\!\!\!\!\!\!\!\!\!
\end{align}
in the ferromagnetic case, with $J^{\text{s}}_{\qq}$, $J^{\text{d}}_{\qq}$ and $J_{\qq}$ defined in the main text,
and $\gamma^{\pm}=1\pm \gamma$.
\\

For the numerical evaluation  of Eqs. \eqref{Eq:AFM_Matrix} and \eqref{Eq:FM_Matrix}, we use a linear system size of $L=800$ and periodic boundary conditions. The interaction $J_{ij} = \pm|\rr_{ij}|^{-\alpha}$ with periodic boundary conditions takes the form
\begin{align}
	J_{ij}=\left(\frac{\sin \left(\frac{\pi}{L}\right)}{\sqrt{\sin(\frac{i_x\pi}{ L})^{2}+\sin(\frac{i_y\pi }{ L})^{2}+\sin(\frac{i_z\pi }{ L})^{2}}}\right)^{\alpha}
\end{align}with $i_x,i_y,i_z\in[0,...,L-1]$

\section{Non-linear sigma model
\label{App:2}}

In this appendix, we give some details on the non-linear sigma model analysis of the antiferromagnetic Heisenberg model. 
We consider the following mapping of a classical spin of length $S$:
\begin{equation}
	\label{eq:Haldane}
	\Svec_j/S = (-1)^j\sqrt{1-\mm_{j}^{2}} \bm{\phi}_{j}+\mm_{j}, 
\end{equation}
where $\ff_j$ describe the order parameter associated with the N\'{e}el state, and $\mm_j$ is the canting field, and they satisfy the conditions $\mm_{j} \cdot \boldsymbol{\phi}_{j}=0$ and $\left|\boldsymbol{\phi}_{j}\right|^{2}=1$, as a consequence of the spin-length conservation. 
Assuming the canting fluctuations to be small, i.e.,  $\mm_j^2 \ll 1$, and by performing a spatial coarse-graining, the effective action of the Heisenberg model reads:
\begin{equation}
	\label{eq:action}
	\mathcal{A}[\ff,\mm] = \int_{t,\rr} \left[S \mm \cdot(\dot{\boldsymbol{\phi}} \times \boldsymbol{\phi})
	- \mathcal{H}(\ff,\mm)
	\right] 
\end{equation}
with $\int_{t,\rr} \equiv \int \dd t\,  \dd^d r$, $\ff = \ff (t,\rr)$, $\mm = \mm(t,\rr)$, and $\mathcal{H}(\ff,\mm)$ is the coarse-grained Hamiltonian associated with Eq.~\ref{eq:Heisenberg_Hamiltonian}. This Hamiltonian density can be obtained in the following way. After replacing the mapping~\eqref{eq:Haldane} into Eq.~\ref{eq:Heisenberg_Hamiltonian}, we retain only the quadratic terms, neglecting higher order processes. This yields:

\begin{multline}
	\label{eq:NLSM-quadratic}
	H \simeq S^2\sum_{i,j} J_{ij}\bigg\{ \big[ (-1)^i \ff_i +\mm_i \big]\cdot\big[ (-1)^j \ff_j +\mm_j \big]\\ 
	+ (\gamma-1)\big[(-1)^i \phi^z_i +m^z_i \big]\big[ (-1)^j \phi^z_j +m^z_j \big] - (-1)^{i+j}\mm_i^2\bigg\}
\end{multline}
where the last term arises from
\begin{equation}
	\sqrt{1-\mm_{i}^{2}} \bm{\phi}_{i} \cdot \sqrt{1-\mm_{j}^{2}} \bm{\phi}_{j}
	\approx 1-\mm_{i}^{2}
\end{equation}
where we approximated $\ff_i \approx \ff_j $ and $\mm_i \approx \mm_j$, as they are slowly varying fields, and used the fact that  $\ff_i^2=1$.
In order to derive the coarse-grained version of the previous Hamiltonian, it is convenient to analyze every term in momentum space, keeping in mind that $\ff$ and $\mm$ are slowly varying fields, i.e., their Fourier component are concentrated around $\qq=0$.
Accordingly, the first term reads
\begin{equation}
\int_\qq J_{\qq+\mathbf{Q}}\,  |\ff_\qq|^2 \simeq
\int_\qq \left(a_0 + a_1{\mathbf{Q}}\cdot \qq + a_2 |\qq|^2 \right) |\ff_\qq|^2,
\end{equation}
with  $\int_{\qq} \equiv 1/(2\pi)^d\int   \dd^d q$, $\mathbf{Q} = (\pi,\cdots,\pi)$ and $|\ff_\qq|^2= \ff_\qq\cdot\ff_{-\qq}$: the first term ($a_0$) gives rise to an inconsequential constant, due to the constraint $|\ff_i|^2=1$, while the second one vanishes since the integrand is odd. Accordingly, the leading term in real space is given by $a_2 |\nabla \ff|^2$, regardless of the form of $J$. A similar analysis can be done for the cross product in Eq.~\eqref{eq:NLSM-quadratic}: the leading order vanishes because of the constraint, ad the remaining terms are not relevant enough to contribute. Finally, the last term in Eq.~\eqref{eq:NLSM-quadratic} reads $\int_\qq J_\qq |m_\qq|^2$, and therefore it depends crucially on the form of $J$. Putting all together, we thus obtain the following Hamiltonian:
\begin{align}
\label{eq:Hamiltonian_quadratic}
H =& \int_\rr \left\{ a_2\, |\nabla \ff|^2 + (\gamma - 1) [a_0 + a_2 (\nabla \phi_z)^2+J_\mathbf{Q}\mm^2]\right\} \nonumber\\
&+ \int_{\rr_1,\rr_2}\!\!\!\!\!\!J_{\rr_1-\rr_2}[ \mm_{\rr_1} \cdot\mm_{\rr_2}+(\gamma-1)m_{z,\rr_1} m_{z,\rr_2}],
\end{align}
with $J(\rr)$ the continuum version of $J_{ij}$. In order to analyse the spectrum of this effective theory, it is convenient to simplify the Berry-phase term in Eq.~\eqref{eq:action}. This can be achieved by using the Ansatz $\ff(\rr) = \ff_0 + \delta\ff(\rr)$, with $\ff_0 = \mathbf{e}_y$ the homogeneous order parameter and $\delta\ff(\rr)$ the fluctuations around it. Since no finite homogeneous solution exist for the canting field, it only consists of small fluctuations, thus we define $\mm(\rr) = \delta\mm(\rr)$. We thus retain only the quadratic terms of the Berry-phase term in Eq.~\eqref{eq:action}, i.e.,
\begin{equation}
\mm \cdot  (\dot{\ff} \times \ff) \simeq \delta m_z\delta \dot{\phi_x} - \delta m_x\delta \dot{\phi_z}.
\end{equation}
By replacing it in the action~\eqref{eq:action}, and considering Eq.~\eqref{eq:Hamiltonian_quadratic}, one realizes that $\delta m_y$ and $\delta \phi_y$ decouple from the remaining degrees of freedom and therefore can be dropped in the following analysis. By defining the multiplet $\bm{\Psi}^T = (\delta \phi_x, \delta \phi_z,\delta m_x, \delta m_z)$, the final quadratic action reads:
\begin{equation}
\mathcal{A}[\bm{\Psi}] = -S^2 \int_{\omega,\qq} \bm{\Psi}^T(-\omega, -\qq) A(\omega, \qq) \bm{\Psi}(\omega, \qq), 
\end{equation}
with the matrix $A(\omega,\qq)$ given by

\begin{equation}
A  = 
\begin{pmatrix}
	a_2 |\qq|^2 & 0 & 0 & -i \frac{\omega}{2S} \\
	0 &  a_0(1-\gamma)\!+\!\gamma a_2|\qq|^2 &   i\frac{\omega}{2S} & 0 \\
	0 &- i  \frac{\omega}{2S} & J_\qq + J_\mathbf{Q} & 0 \\
	i  \frac{\omega}{2S} & 0 & 0 & \gamma J_\qq + J_\mathbf{Q}
\end{pmatrix}.
\end{equation}
The excitation spectrum $E_\qq$ can be finally obtained by solving the equation $\text{det} A(E_\qq,\qq) = 0$, and leads to:
\begin{subequations}
\label{anti_spec}
\begin{align}
	E^+_\qq & = 2S\sqrt{a_2} |\qq| \sqrt{J_\mathbf{Q}+\gamma J_\qq}, \\
	E^-_\qq & = 2S\sqrt{(1-\gamma)a_0+\gamma a_2|\qq|^2}  \sqrt{J_\mathbf{Q}+J_\qq},
\end{align}
\end{subequations}
where $E^+_\qq$ corresponds to the Goldstone branch.

\section{Regularization of super-extensive divergences}
\label{app_c}
\begin{figure*}[t!]
\includegraphics[width=\textwidth]{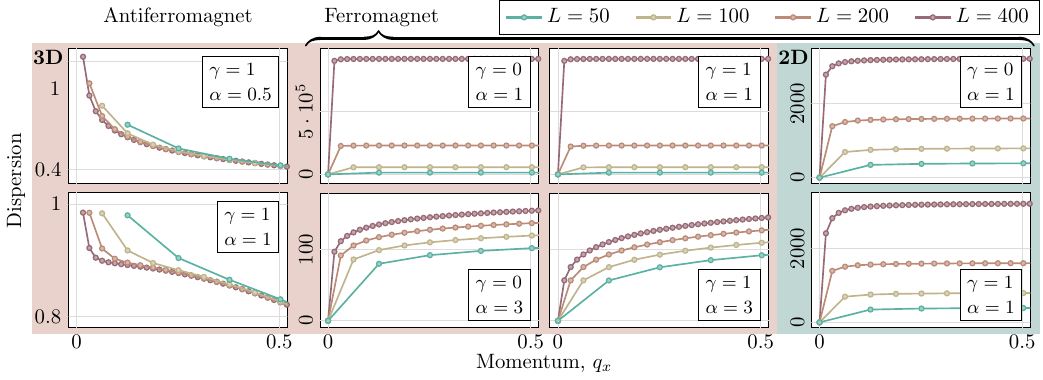}
\caption{Unrescaled spin wave dispersions 
	for different linear system sizes $L$. Their rescaled form is shown in the insets of Fig.~\ref{fig:Dispersions}. }
\label{fig:DispersionsSM}
\end{figure*}
Here, we inspect the origin of the divergence in the antiferromagnetic excitation spectrum in Eq.\,\eqref{anti_spec} and discuss its physical interpretation. Indeed, in analogy with the ferromagnetic case, the divergence of the long-range interacting contribution in the last term of Hamiltonian~\eqref{eq:Hamiltonian_quadratic} causes the appearance of a super-extensive contribution to the system internal energy.
This is easily seen, as the integral $J_\qq = \int_\rr J(\rr) \ee^{i\qq\cdot\rr}$ diverges with the system size $L^d$ for $\alpha \leq d-2$.
Here, super-extensivity occurs since the long-range contribution of the canting field, second term on the r.h.s. of Eq.\,\eqref{eq:Hamiltonian_quadratic}, is convolved with the quadratic dispersion generated by the order parameter, first term  on the r.h.s. of Eq.\,\eqref{eq:Hamiltonian_quadratic}.

Thus, for $\alpha \leq d-2$, the frequency of low-energy modes increases with the size of the system $E_\qq \sim L^{(d-2-\alpha)/2}$. Indeed, long-range interactions are well known in many cases to cause a scaling of the relevant time scales with system size~\cite{antoni1995clustering}. In particular, for long-range quantum lattice models the fastest time scale $\tau$ was found to shrink algebraically with an increasing system size $N$, i.e., $\tau\propto N^{-p}$, where $p$ is a positive exponent. It follows that  the excitations propagate increasingly faster as the thermodynamic limit is approached and, hence, their velocity is not bounded by any finite quantity in the absolute time $t$.  

Long-range interactions have the additional effect of rendering the spacing between the Hamiltonian low-energy eigenvalues (i.e., the quasi-particle spectrum) finite even in the thermodynamic limit: in other words, the Hamiltonian spectrum remains point-like even for $L\to \infty$. To show this, we consider a $d$-dimensional system with periodic boundary conditions and finite linear size $L$. The momenta are then quantized according to $\qq = \mathbf{n} 2\pi/L$, with the $d$-dimensional vector $\mathbf{n}\equiv (n_1,\dots,n_d)$ with $n_i$, $i=1,\dots,d$ integer numbers. The interaction also obeys periodic boundary conditions as $J(\rr) = d(\rr/L)^{-1} (\pi/L)^\alpha$, where
\begin{equation}
d(\rr/L) = \left[\sum_{i=1}^d\sin\left(\frac{\pi x_i}{L}\right)^2\right]^\frac{\alpha}{2}\!,
\end{equation}
with $\rr=(x_1,\dots,x_d)$ the $d$-dimensional spatial vector. The Fourier transform of $J(\rr)$ can the be written as
\begin{equation}
J(\qq)=L^{d-\alpha} \pi^\alpha \int_{\bar{\rr}} \frac{\ee^{i 2\pi \mathbf{n}\cdot\bar{\rr}}}{d(\bar{\rr})},
\end{equation}
with $\int_{\bar{\rr}}\equiv \int_0^1 \dd \bar{x}_1\dots\dd \bar{x}_d$, upon the substitution $\rr = L \bar{\rr}$ in the integral. From the dispersion $E_\qq = |\qq|\sqrt{J_\qq}$ one then finds
\begin{equation}
\label{eq:rescaled_energy}
E_\mathbf{n} = L^{\frac{d-\alpha-2}{2}} 2\pi^{\frac{2+\alpha}{2}}|\mathbf{n}|\sqrt{\int_{\bar{\rr}} \frac{\ee^{i 2\pi \mathbf{n}\cdot\bar{\rr}}}{d(\bar{\rr})}} \equiv L^{\frac{d-\alpha-2}{2}} \epsilon_\mathbf{n},
\end{equation}
where $\epsilon_\mathbf{n}$ is a regular, $L$-independent function of the quantum number $\mathbf{n}$. We then define the spectral spacing as $\Delta_\mathbf{n} \equiv E_{\mathbf{n+e}}-E_{\mathbf{n}}$, with $\mathbf{e}$ a unit vector in an arbitrary direction. It then follows that 
\begin{equation}
\Delta_\mathbf{n} \propto L^{\frac{d-\alpha-2}{2}},
\end{equation}
entailing that, in the thermodynamic limit $L\to \infty$:
\begin{equation}
\begin{cases}
	\Delta_\mathbf{n} \to 0 & \text{for } \alpha>d-2, \\
	\Delta_\mathbf{n} \to \text{const.} & \text{for } \alpha=d-2, \\
	\Delta_\mathbf{n} \to \infty & \text{for } \alpha<d-2.
\end{cases}
\end{equation}
It then follows that, while for $\alpha> d-s$ the spectrum becomes continuous, for $\alpha\leq d-2$ it remains discrete. The diverging character of $\Delta_\mathbf{n}$, related to the energy super-extensivity mentioned above, will be regularized by the rescaling discussed in the following. 

In analogy with several studies of excitation propagation and dynamics in quantum long-range systems\,\cite{bachelard2013universal,storch2015interplay}, we will introduce the rescaled time $t'= t L^{p}$ and, accordingly, the rescaled frequencies $\omega'L^{-p}$ in order to obtain a finite dispersion relation in the thermodynamic limit. 

In the new variables the coupling matrix reads
\begin{equation}
A  \!=\! 
\begin{pmatrix}
	a_2 |\qq|^2 & 0 & 0 & -i \frac{\omega'L^p}{2S} \\
	0 &  a_0(1-\gamma)\!+\!\gamma a_2|\qq|^2 &   i\frac{\omega'L^p}{2S} & 0 \\
	0 &- i  \frac{\omega'L^p}{2S} & J_\qq & 0 \\
	i  \frac{\omega'L^p}{2S} & 0 & 0 & \gamma J_\qq
\end{pmatrix}
\end{equation}
yielding the rescaled dispersion relation (for $\gamma=1$)
\begin{equation}
\label{eq:disp_AFM}
E_\qq \propto \frac{|\qq|}{L^p} \sqrt{J_\qq} .
\end{equation}
The latter result remains finite for all $\alpha$ as long as we impose
\begin{align}
p=\begin{cases}
	0\,\,\mathrm{if}\,\,\alpha\geq d-2,\\
	\frac{d-2-\alpha}{2}\,\,\mathrm{if}\,\,\alpha<d-2,
\end{cases}
\end{align}
which extends the conventional Kac scaling procedure to the antiferromagnetic regime. 
By applying this rescaling to Eq.~\eqref{eq:rescaled_energy}, one sees that the spectrum for $\alpha \leq d-2$ remains discrete in the thermodynamic limit and it is given by $\epsilon_\mathbf{n}$. In contrast, for $\alpha>d-2$, the spectrum is continuous in the thermodynamic limit and can be expressed in terms of continuous momenta $\qq$.

As a result of the ``antiferromagnetic" Kac rescaling the excitations spectrum of the system remains gapped in the entire regime $\alpha<d-2$ as in the marginal case $\alpha=d-2$.
The entire spectrum becomes discrete in the new variables and the system will develop all the typical out-of-equilibrium features of strong long-range systems, such as quasi-stationary-states\,\cite{kastner2011diverging,bachelard2013universal,Defenu_2021_pnas}, adiabaticity loss\,\cite{defenu2018dynamical}, and several others\,\cite{Defenu_2021}.

\end{document}